\documentclass[twoside, final]{article}%
\usepackage{amssymb}
\usepackage{amsfonts}
\usepackage{amsmath}
\usepackage{graphicx}%
\providecommand{\U}[1]{\protect\rule{.1in}{.1in}}
\newtheorem{theorem}{Theorem}

\newtheorem{corollary}[theorem]{Corollary}

\newtheorem{lemma}[theorem]{Lemma}

\newenvironment{proof}[1][Proof]{\noindent\textbf{#1.} }{\ \rule{0.5em}{0.5em}}
\begin{document}

\title{\textbf{Perturbational Blowup Solutions to the 2-Component Camassa-Holm
Equations}}
\author{M\textsc{anwai Yuen\thanks{E-mail address: nevetsyuen@hotmail.com }}\\\textit{Department of Applied Mathematics, The Hong Kong Polytechnic
University,}\\\textit{Hung Hom, Kowloon, Hong Kong}}
\date{Revised 08-Dec-2010}
\maketitle

\begin{abstract}
In this article, we study the perturbational method to construct the
non-radially symmetric solutions of the compressible 2-component Camassa-Holm
equations. In detail, we first combine the substitutional method and the
separation method to construct a new class of analytical solutions for that
system. In fact, we perturb the linear velocity:%
\begin{equation}
u=c(t)x+b(t),
\end{equation}
and substitute it into the system. Then, by comparing the coefficients of the
polynomial, we can deduce the functional differential equations involving
$(c(t),b(t),\rho^{2}(0,t)).$ Additionally, we could apply the Hubble's
transformation
\begin{equation}
c(t)=\frac{\dot{a}(3t)}{a(3t)},
\end{equation}
to simplify the ordinary differential system involving $(a(3t),b(t),\rho
^{2}(0,t))$. After proving the global or local existences of the corresponding
dynamical system, a new class of analytical solutions is shown. And the
corresponding solutions in radial symmetry are also given. To determine that
the solutions exist globally or blow up, we just use the qualitative
properties about the well-known Emden equation:%
\begin{equation}
\left\{
\begin{array}
[c]{c}%
\frac{d^{2}}{dt^{2}}a(3t)=\frac{\xi}{a^{\frac{1}{3}}(3t)}\text{,}\\
a(0)=a_{0}>0\text{, }\dot{a}(0)=a_{1}%
\end{array}
\right.
\end{equation}
Our solutions obtained by the perturbational method, fully cover the previous
known results in "M.W. Yuen, \textit{Self-Similar Blowup Solutions to the
2-Component Camassa-Holm Equations, }J. Math. Phys., \textbf{51} (2010)
093524, 14pp." by the separation method.



\end{abstract}

\section{Introduction}

The 2-component Camassa-Holm equations of shallow water system can be
expressed by%
\begin{equation}
\left\{
\begin{array}
[c]{c}%
\rho_{t}+u\rho_{x}+\rho u_{x}=0,\text{ }x\in R\\
m_{t}+2u_{x}m+um_{x}+\sigma\rho\rho_{x}=0
\end{array}
\right.  \label{2com}%
\end{equation}
with
\begin{equation}
m=u-\alpha^{2}u_{xx}. \label{meq}%
\end{equation}
Here $u=u(x,t)\in R$ and $\rho=\rho(x,t)\geq0$ are the velocity and the
density of fluid respectively. The constant $\sigma$ is equal to $1$ or $-1.$
If $\sigma=-1,$ the gravity acceleration points upwards \cite{CLZ}, \cite{C1},
\cite{GY}, \cite{GZ} and \cite{G}. For $\sigma=1,$ the researches regarding
the corresponding models could be referred by \cite{CI}, \cite{ELY2},
\cite{GZ} and \cite{GY}. When $\rho\equiv0,$ the system returns to the
Camassa-Holm equation \cite{CH}. The searching of Camassa-Holm equation can
capture breaking waves. Peaked traveling waves is a long-standing open problem
\cite{W}.

In 2010, Yuen used the separation method to obtain a class of blowup or global
solutions of the Camassa-Holm equations \cite{Yuen 3} and Degasperis-Procesi
equations \cite{Yuen 4}. In particular, for the integrable system of the
Camassa-Holm equations with $\sigma=1$, we have the global solutions:%
\begin{equation}
\left\{
\begin{array}
[c]{c}%
\rho(x,t)=\max\left\{  \frac{f\left(  \eta\right)  }{a(3t)^{1/3}},0\right\}
,\text{ }u(x,t)=\frac{\overset{\cdot}{a}(3t)}{a(3t)}x\\
\overset{\cdot\cdot}{a}(s)-\frac{\xi}{3a(s)^{1/3}}=0,\text{ }a(0)=a_{0}%
>0,\text{ }\overset{\cdot}{a}(0)=a_{1}\\
f(\eta)=\xi\sqrt{-\frac{\eta^{2}}{\xi}+\left(  \xi\alpha\right)  ^{2}}%
\end{array}
\right.  \label{caca}%
\end{equation}
where $\eta=\frac{x}{a(s)^{1/3}}$ with $s=3t;$ $\xi>0$ and $\alpha\geq0$ are
arbitrary constants. \cite{Yuen 3}

Meanwhile, the isentropic compressible Euler equations can be written in the
following form:
\begin{equation}
\left\{
\begin{array}
[c]{rl}%
{\normalsize \rho}_{t}{\normalsize +\nabla\cdot\rho u} & {\normalsize =}%
{\normalsize 0}\\
{\normalsize (\rho u)}_{t}{\normalsize +\nabla\cdot(\rho u\otimes u)+\nabla P}
& {\normalsize =0.}%
\end{array}
\right.  \label{Euler}%
\end{equation}
As usual, $\rho=\rho(x,t)$ and $u=u(x,t)\in\mathbf{R}^{N}$ are the density and
the velocity respectively with $x=(x_{1},$ $x_{2},$ $...,$ $x_{N})\in R^{N}$.
For some fixed $K>0$, we have a $\gamma$-law on the pressure $P=P(\rho)$, i.e.%
\begin{equation}
{\normalsize P}\left(  \rho\right)  {\normalsize =K\rho}^{\gamma}
\label{gamma}%
\end{equation}
with a constant $\gamma\geq1$. For solutions in radially symmetry:%
\begin{equation}
\rho(x,t)=\rho(r,t)\text{ and }u(x,t)=\frac{x}{r}V(r,t)=:\frac{x}{r}V
\end{equation}
where the radial $r=\sum_{i=1}^{N}x_{i}^{2}$,\newline the compressible Euler
equations (\ref{Euler}) become,%
\begin{equation}
\left\{
\begin{array}
[c]{rl}%
\rho_{t}+V\rho_{r}+\rho V_{r}+{\normalsize \frac{N}{r}\rho V} &
{\normalsize =0}\\
\rho\left(  V_{t}+VV_{r}\right)  +\nabla P & {\normalsize =0.}%
\end{array}
\right.  \label{gamma=1}%
\end{equation}
Recently, there are some researches concerning the construction of solutions
of the compressible Euler and Navier-Stokes equations by the substitutional
method \cite{Li}, \cite{LW}, \cite{Y1} and \cite{Liang}. They assume that the
velocity is linear:%
\begin{equation}
u(x,t)=c(t)x
\end{equation}
and substitute it into the system to derive the dynamic system about the
function $c(t)$. Then they use the standard argument of phase diagram to drive
the blowup or global existence of the ordinary differential equation involving
$c(t)$.\newline On the other hand, the separation method can be governed to
seek the radial symmetric solutions by the functional form:%
\begin{equation}
\rho(r,t)=\frac{f(\frac{r}{a(t)})}{a^{N}(t)}\text{ and }V(r,t)=\frac{\dot
{a}(t)}{a(t)}r.
\end{equation}
(\cite{GW}, \cite{M1}, \cite{DXY1}, \cite{Li}, \cite{Y1}, \cite{Y44},
\cite{Y2} and \cite{Y3})

It is natural to consider the more general linear velocity:%
\begin{equation}
u(x,t)=c(t)x+b(t) \label{linearv}%
\end{equation}
to construct new solutions. In this article, we can first combine the two
conventional approaches (substitutional method and separation method) to
derive the corresponding solutions for the system. In fact, the main theme of
this article is to substitute the linear velocity (\ref{linearv}) into the
Camassa-Holm equations (\ref{2com}) and compare the coefficient of the
different polynomial degrees for deducing the functional differential
equations involving $(c(t),b(t),\rho^{2}(0,t)).$ Then, we can apply the
Hubble's transformation
\begin{equation}
c(t)=\frac{\dot{a}(3t)}{a(3t)}%
\end{equation}
with $\dot{a}(3t):=\frac{da(3t)}{dt}$, to simplify the functional differential
system involving $(a(3t),b(t),\rho^{2}(0,t))$. After proving the local
existences of the corresponding dynamical system, we can show the results below:

\begin{theorem}
\label{thm:1 copy(1)}For the 2-component Camassa-Holm equations (\ref{2com}),
there exists a family of solutions,
\begin{equation}
\left\{
\begin{array}
[c]{c}%
\rho^{2}(x,t)=\max\left\{  \rho^{2}(0,t)-\frac{2}{\sigma}[\dot{b}%
(t)+3b(t)\frac{\dot{a}(t)}{a(t)}]x-\frac{3\xi}{\sigma a^{\frac{4}{3}}(t)}%
x^{2},\text{ }0\right\} \\
{\normalsize u(x,t)=}\frac{\overset{\cdot}{a}(3t)}{a(3t)}x+b(t)\\
\frac{d^{2}}{dt^{2}}a(3t)=\frac{\xi}{a^{\frac{1}{3}}(3t)},\text{ }%
a(0)=a_{0}>0\text{, }\dot{a}(0)=a_{1}\\
\frac{d^{2}}{dt^{2}}b(t)+\frac{6\dot{a}(3t)}{a(3t)}\frac{d}{dt}b(t)+\frac
{12\xi}{a^{\frac{4}{3}}(3t)}b(t)=0,b(0)=b_{0}\text{, }\dot{b}(0)=b_{1}\\
\frac{d}{dt}\left[  \rho^{2}(0,t)\right]  =\rho^{2}(0,t)-\frac{2}{\sigma}%
[\dot{b}(t)+3b(t)\frac{\dot{a}(3t)}{a(3t)}]x-\frac{3\xi}{\sigma a^{2}%
(3t)}x^{2},\text{ }\rho^{2}(0,0)=\alpha^{2}\text{ }%
\end{array}
\right.  \label{solutionper}%
\end{equation}
where $a_{0}$, $a_{1}$, $b_{1}$, $b_{2}$ and $\alpha$ are arbitrary constants.
\end{theorem}

We remark that the above solutions (\ref{solutionper}) fully cover the
previous known results \cite{Yuen 3} by the separation method by choosing
$b_{0}=b_{1}=0$.

\section{Perturbational Method}

The proof for Theorem \ref{thm:1 copy(1)} requires the standard manipulation
of algebraic computation only:

\begin{proof}
The momentum equation (\ref{2com})$_{2}$, becomes%
\begin{equation}
\left(  u-u_{xx}\right)  _{t}+2u_{x}(u-u_{xx})+u(u-u_{xx})_{x}+\sigma\rho
\rho_{x}=0. \label{abc}%
\end{equation}
First, we perturb the velocity with this following functional form:
\begin{equation}
u(x,t)=c(t)x+b(t) \label{linearvelocity}%
\end{equation}
where $b(t)$ and $c(t)$ are the time functions determined later.

As the velocity $u$ (\ref{linearvelocity}) is linear:
\begin{equation}
u_{xx}=0,
\end{equation}
it can be simplified to be%
\begin{equation}
u_{t}+3uu_{x}+\sigma\rho\rho_{x}=0
\end{equation}%
\begin{equation}
\dot{c}(t)x+\dot{b}(t)+3[c(t)x+b(t)]c(t)+\frac{\sigma}{2}\frac{\partial
}{\partial x}\rho^{2}=0
\end{equation}%
\begin{equation}
\frac{\sigma}{2}\frac{\partial}{\partial x}\rho^{2}=-[\dot{b}%
(t)+3b(t)c(t)]-[\dot{c}(t)+3c^{2}(t)]x.
\end{equation}
Then, we take integration from $[0,x]$ to have:%
\begin{equation}
\frac{\sigma}{2}\int_{0}^{x}\frac{\partial}{\partial s}\rho^{2}ds=-[\dot
{b}(t)+3b(t)c(t)]\int_{0}^{x}ds-[\dot{c}(t)+3c^{2}(t)]\int_{0}^{x}sds
\label{eq123}%
\end{equation}%
\begin{equation}
\frac{\sigma}{2}\left[  \rho^{2}(x,t)-\rho^{2}(0,t)\right]  =-[\dot
{b}(t)+3b(t)c(t)]x-\frac{[\dot{c}(t)+3c^{2}(t)]}{2}x^{2}%
\end{equation}%
\begin{equation}
\rho^{2}(x,t)=\rho^{2}(0,t)-\frac{2}{\sigma}[\dot{b}(t)+3b(t)c(t)]x-\frac
{[\dot{c}(t)+3c^{2}(t)]}{\sigma}x^{2}. \label{density}%
\end{equation}
On the other hand, for the 1-dimensional mass equation (\ref{2com})$_{1}$, we
obtain%
\begin{equation}
\rho_{t}+\left[  c(t)x+b(t)\right]  \rho_{x}+\rho c(t)=0.
\end{equation}
Here, we multiple $\rho$ on both sides to have%
\begin{equation}
\frac{1}{2}\left(  \rho^{2}\right)  _{t}+\frac{\left[  c(t)x+b(t)\right]  }%
{2}\left(  \rho^{2}\right)  _{x}+\rho^{2}c(t)=0. \label{massmass}%
\end{equation}
After that, we can substitute equation (\ref{density}) into equation
(\ref{massmass}):%
\begin{align}
&  \frac{1}{2}\left(  \frac{\partial}{\partial t}\left[  \rho^{2}(0,t)\right]
-\frac{2}{\sigma}\frac{\partial}{\partial t}[\dot{b}(t)+3b(t)c(t)]x-\frac
{\partial}{\partial t}\frac{[\dot{c}(t)+3c^{2}(t)]}{\sigma}x^{2}\right)
\\[0.1in]
&  +\left[  c(t)x+b(t)\right]  \left(  -\frac{1}{\sigma}[\dot{b}%
(t)+3b(t)c(t)]-\frac{1}{\sigma}[\dot{c}(t)+3c^{2}(t)]x\right) \\[0.1in]
&  +c(t)\left[  \rho^{2}(0,t)-\frac{2}{\sigma}[\dot{b}(t)+3b(t)c(t)]x-\frac
{[\dot{c}(t)+3c^{2}(t)]}{\sigma}x^{2}\right]
\end{align}%
\begin{align}
&  =\frac{1}{2}\frac{\partial}{\partial t}\left[  \rho^{2}(0,t)\right]
+c(t)\rho^{2}(0,t)-\frac{b(t)}{\sigma}[\dot{b}(t)+3b(t)c(t)]\\[0.1in]
&  +\left\{
\begin{array}
[c]{c}%
-\frac{1}{\sigma}\frac{\partial}{\partial t}[\dot{b}(t)+3b(t)c(t)]-\frac
{c(t)}{\sigma}[\dot{b}(t)+3b(t)c(t)]\\
-\frac{b(t)}{\sigma}[\dot{c}(t)+3c^{2}(t)]-\frac{2c(t)}{\sigma}[\dot
{b}(t)+3b(t)c(t)]
\end{array}
\right\}  x\\[0.1in]
&  +\left\{
\begin{array}
[c]{c}%
-\frac{1}{2\sigma}\frac{\partial}{\partial t}[\dot{c}(t)+3c^{2}(t)]-\frac
{1}{\sigma}[\dot{c}(t)+3c^{2}(t)]c(t)\\
-\frac{c(t)[\dot{c}(t)+3c^{2}(t)]}{\sigma}%
\end{array}
\right\}  x^{2}%
\end{align}
By comparing the coefficients of the polynomial, we require the functional
differential equations involving $(c(t),b(t),\rho^{2}(0,t))$:%
\begin{equation}
\left\{
\begin{array}
[c]{c}%
\frac{d}{dt}\left[  \rho^{2}(0,t)\right]  +2c(t)\rho^{2}(0,t)-\frac{2}{\sigma
}b(t)[\dot{b}(t)+3b(t)c(t)]=0\\
\frac{d}{dt}[\dot{b}(t)+3b(t)c(t)]+3c(t)[\dot{b}(t)+3b(t)c(t)]+b(t)[\dot
{c}(t)+3c^{2}(t)]=0\\
\frac{d}{dt}[\dot{c}(t)+3c^{2}(t)]+4[\dot{c}(t)+3c^{2}(t)]c(t)=0
\end{array}
\right.  \label{ODE}%
\end{equation}
For details (existence, uniqueness and continuous dependence) about general
functional differential equations, the interested reader may refer to the
classical literatures \cite{H} and \cite{Wa}.

For solving the above ordinary differential system (\ref{ODE}), we initially
solve equation (\ref{ODE})$_{3}$ about the function $c(t)$. Here we let the
function $c(t)$ be expressed with the Hubble's transformation:%
\begin{equation}
c(t)=\frac{\dot{a}(3t)}{a(3t)}%
\end{equation}
where $\dot{a}(3t):=\frac{da(3t)}{dt}$ and the function $a(3t)$ could be
determined later.\newline It is transformed to be%
\begin{equation}
\frac{d}{dt}[\frac{3\ddot{a}(3t)}{a(3t)}-\frac{3\dot{a}^{2}(3t)}{a^{2}%
(3t)}+\frac{3\dot{a}^{2}(3t)}{a^{2}(3t)}]+4[\frac{3\ddot{a}(3t)}{a(3t)}%
-\frac{3\dot{a}^{2}(3t)}{a^{2}(3t)}+\frac{3\dot{a}^{2}(3t)}{a^{2}(3t)}%
]\frac{\dot{a}(3t)}{a(3t)}=0
\end{equation}%
\begin{equation}
\left\{
\begin{array}
[c]{c}%
\frac{d}{dt}\left(  \frac{\ddot{a}(3t)}{a(3t)}\right)  +\frac{4\ddot{a}%
(3t)}{a(3t)}\frac{\dot{a}(3t)}{a(3t)}=0\\
a(0)=a_{0}>0,\text{ }\dot{a}(0)=a_{1}\text{, }\ddot{a}(0)=a_{2}%
\end{array}
\right.
\end{equation}%
\begin{equation}
\frac{3\dddot{a}(3t)}{a(3t)}-\frac{3\dot{a}(3t)\ddot{a}(3t)}{a^{2}(3t)}%
+\frac{4\dot{a}(3t)\ddot{a}(3t)}{a^{2}(3t)}=0
\end{equation}%
\begin{equation}
\frac{\dddot{a}(3t)}{a(3t)}+\frac{\dot{a}(3t)\ddot{a}(3t)}{3a^{2}(3t)}=0.
\end{equation}
Then, we multiple $a^{2}(3t)$ on both sides to have:%
\begin{equation}
a(3t)\dddot{a}(3t)+\frac{\dot{a}(3t)\ddot{a}(3t)}{3}=0.
\end{equation}
It can be reduced to the second-order Emden equation:%
\begin{equation}
\left\{
\begin{array}
[c]{c}%
\frac{d^{2}}{dt^{2}}a(3t)=\frac{\xi}{a^{\frac{1}{3}}(3t)}\\
a(0)=a_{0}>0\text{, }\dot{a}(0)=a_{1}%
\end{array}
\right.  \label{Emdeneq1}%
\end{equation}
where $\xi:=a_{0}^{\frac{1}{3}}a_{2}$ is an arbitrary constant by choosing
$a_{2}.$\newline We remark that the well-known Emden equation is well studied
in astrophysics and mathematics.

Next, for the second equation (\ref{ODE})$_{2}$ about $b(t)$ of the functional
differential system, we could further simply it in terms of the known function
$a(3t)$:%
\begin{equation}
\frac{d}{dt}[\dot{b}(t)+3b(t)\frac{\dot{a}(3t)}{a(3t)}]+3\frac{\dot{a}%
(3t)}{a(3t)}[\dot{b}(t)+3b(t)\frac{\dot{a}(3t)}{a(3t)}]+\frac{3\ddot{a}%
(3t)}{a(3t)}b(t)=0
\end{equation}%
\begin{equation}
\ddot{b}(t)+6\frac{\dot{a}(3t)}{a(3t)}\dot{b}(t)+\left[  9\frac{\ddot{a}%
(3t)}{a(3t)}-9\frac{\dot{a}^{2}(3t)}{a^{2}(3t)}+\frac{9\dot{a}^{2}(3t)}%
{a^{2}(3t)}+\frac{3\ddot{a}(3t)}{a(3t)}\right]  b(t)=0
\end{equation}%
\begin{equation}
\left\{
\begin{array}
[c]{c}%
\ddot{b}(t)+6\frac{\dot{a}(3t)}{a(3t)}\dot{b}(t)+\frac{12\xi}{a^{\frac{4}{3}%
}(3t)}b(t)=0,\\
b(0)=b_{0}\text{, }\dot{b}(0)=b_{1}%
\end{array}
\right.
\end{equation}
with the Emden equation (\ref{Emdeneq1}).\newline We denote $f_{1}%
(t)=6\frac{\dot{a}(3t)}{a(3t)}$ and $f_{2}(t)=\frac{12\xi}{a^{\frac{4}{3}%
}(3t)}$ to have%
\begin{equation}
\left\{
\begin{array}
[c]{c}%
\ddot{b}(t)+f_{1}(t)\dot{b}(t)+f_{2}(t)b(t)=0\\
b(0)=b_{0}\text{, }\dot{b}(0)=b_{1}.
\end{array}
\right.
\end{equation}
Therefore, when the functions $f_{1}(t)$ and $f_{2}(t)$ are bounded, that is
\begin{equation}
\left\vert f_{1}(t)\right\vert \leq F_{1}\text{ and }\left\vert f_{2}%
(t)\right\vert \leq F_{2}%
\end{equation}
with some constants $F_{1}$ and $F_{2}$, provided that the functions $\frac
{1}{a(3t)}$ and $\dot{a}(3t)$ exist, the functions $b(t)$ and $\dot{b}(t)$ can
be guaranteed for existing by the comparison theorem of ordinary differential
equations \cite{W}.\newline Lastly, for the first equation (\ref{ODE})$_{1}$,
we denote $H(t)=\frac{2\dot{a}(3t)}{a(3t)}$ and $G(t)=\frac{2b(t)}{\sigma
}[\dot{b}(t)+3b(t)\frac{\dot{a}(3t)}{a(3t)}]$ in terms of functions $\frac
{1}{a(3t)},$ $a(3t)$, $b(t)$ and $\dot{b}(t)$ provided that they exists, to
solve%
\begin{equation}
\left\{
\begin{array}
[c]{c}%
\frac{d}{dt}\left[  \rho^{2}(0,t)\right]  +\rho^{2}(0,t)H(t)=G(t)\\
\rho^{2}(0,0)=\alpha^{2}.
\end{array}
\right.  \text{ } \label{eqeqr}%
\end{equation}
The formula of the first-order ordinary differential equation (\ref{eqeqr}) is%
\begin{equation}
\rho^{2}(0,t)=\frac{\int_{0}^{t}\mu(s)G(s)ds+k}{\mu(t)}%
\end{equation}
where%
\begin{equation}
\mu(t)=e^{\int_{0}^{t}H(s)ds}.
\end{equation}
Therefore, we have the density function from equation (\ref{density}):%
\begin{equation}
\rho^{2}(x,t)=\rho^{2}(0,t)-\frac{2}{\sigma}[\dot{b}(t)+3b(t)\frac{\dot
{a}(3t)}{a(3t)}]x-\frac{3\xi}{\sigma a^{\frac{4}{3}}(3t)}x^{2}.
\end{equation}

For $\rho(x,t)\geq0$, we may set%
\begin{equation}
\rho^{2}(x,t)=\max\left\{  \rho^{2}(0,t)-\frac{2}{\sigma}[\dot{b}%
(t)+3b(t)\frac{\dot{a}(3t)}{a(3t)}]x-\frac{3\xi}{\sigma a^{\frac{4}{3}}%
(3t)}x^{2},\text{ }0\right\}  . \label{solutions88}%
\end{equation}
In conclusion, we have the corresponding functional differential equations
(\ref{solutionper}) to be the solutions of Camassa-Holm equations.

The proof is completed.
\end{proof}

We notice that the above solutions are not radially symmetric for the density
function $\rho$ with $b(t)\neq0$. Thus, the above solutions, cannot be
obtained by the separation method of the self-similar functional \cite{Yuen
3}, as%
\begin{equation}
\rho(x,t)\neq f(\frac{x}{a(3t)})g(a(3t))\text{ and }u(x,t)=\frac{\dot{a}%
(3t)}{a(3t)}x+b(t).
\end{equation}

On the other hand, for the 2-component Camassa-Holm equations in radial
symmetry with linear velocity $u(r,t)$:%
\begin{equation}
\left\{
\begin{array}
[c]{rl}%
\rho_{t}+V\rho_{r}+\rho V_{r} & {\normalsize =0}\\
V_{t}+3VV_{r}+\sigma\rho\rho_{r} & {\normalsize =0,}%
\end{array}
\right.
\end{equation}
we may replace equation (\ref{eq123}) to have the corresponding step by taking
the integration from $[0,$ $r]$%
\begin{equation}
\frac{\sigma}{2}\int_{0}^{r}\frac{\partial}{\partial s}\rho^{2}ds=-[\dot
{b}(t)+b(t)c(t)]\int_{0}^{r}ds-3[\dot{c}(t)+c^{2}(t)]\int_{0}^{r}sds.
\end{equation}
It is clear for that the rest of proof is similar to have the corresponding
result for the solutions in radial symmetry:

\begin{theorem}
\label{thm:1}For the 2-component Camassa-Holm equations in radial symmetry
(\ref{2com}), there exists a family of solutions,%
\begin{equation}
\left\{
\begin{array}
[c]{c}%
\rho^{2}(r,t)=\max\left\{  \rho^{2}(0,t)-\frac{2}{\sigma}[\dot{b}%
(t)+3b(t)\frac{\dot{a}(t)}{a(t)}]r-\frac{3\xi}{\sigma a^{\frac{4}{3}}(t)}%
r^{2},\text{ }0\right\} \\
{\normalsize u(r,t)=}\frac{\overset{\cdot}{a}(3t)}{a(3t)}r+b(t)\\
\frac{d}{dt}\left[  \rho^{2}(0,t)\right]  =\rho^{2}(0,t)-\frac{2}{\sigma}%
[\dot{b}(t)+3b(t)\frac{\dot{a}(3t)}{a(3t)}]r-\frac{3\xi}{\sigma a^{2}%
(3t)}r^{2}\text{, }\rho^{2}(0,0)=\alpha^{2}%
\end{array}
\right.  \label{radialsy}%
\end{equation}
where $a(3t)$ and $b(t)$ are the solutions of equations (\ref{solutionper}%
)$_{3}$ and (\ref{solutionper})$_{4}$.
\end{theorem}

\section{Blowup or Global Solutions}

To determine that the solutions are global or local only, we can use the
corresponding lemma about the Emden equation:

\begin{lemma}
\label{lemma33 copy(1)}For the Emden equation (\ref{solutionper})$_{3}$,%
\begin{equation}
\left\{
\begin{array}
[c]{c}%
\ddot{a}(3t)=\frac{\xi}{a^{\frac{1}{3}}(3t)}\\
a(0)=a_{0}>0,\text{ }\dot{a}(0)=a_{1},
\end{array}
\right.  \label{emden1/2}%
\end{equation}
(1) if $\xi<0$, there exists a finite time $T$, such that
\begin{equation}
\underset{t\rightarrow T^{-}}{\lim}a(3t)=0.
\end{equation}
(2) if $\xi=0$, with $a_{1}<0$, the solution $a(t)$ blows up in the finite
time:
\begin{equation}
T=\frac{-a_{0}}{a_{1}},
\end{equation}
(3) otherwise, the solution $a(t)$ exists globally.
\end{lemma}

We observe that it is the same lemma for the function $a(s)$ with $s=3t$ by
the separation methods in \cite{Yuen 3}. Therefore, the proofs can be found in
Lemma 3 of \cite{Yuen 3}.\newline The gradient of the velocity in solutions
(\ref{solutionper}) and (\ref{radialsy}), is
\begin{equation}
\frac{\partial}{\partial x}u(x,t)=\frac{\partial}{\partial r}u(r,t)=\frac
{\dot{a}(3t)}{a(3t)}.
\end{equation}
When the function $a(t)$ blows up with a finite time $T$, $\frac{\partial
}{\partial x}u(x,T)$ also blows up at every point $x.$ And based on the above
lemma about the Emden equation for $a(t)$, it is clear to have the corollary below:

\begin{corollary}
(1a) For $\xi<0$, solutions (\ref{solutionper}) and (\ref{radialsy}) blow up
in a finite time $T;$\newline(1b) For $\xi=0$, with $a_{1}<0$, solutions
(\ref{solutionper}) and (\ref{radialsy}) blow up in the finite time:%
\begin{equation}
T=\frac{-a_{0}}{a_{1}}.
\end{equation}
(2) otherwise, solutions (\ref{solutionper}) and (\ref{radialsy}) exist globally.
\end{corollary}


\begin{thebibliography}{99}                                                                                               %
\bibitem {CH}R. Camassa and D. D. Holm, \textit{Intergrable Shallow Water
Equation with Peaked Solitons}, Phys. Rev. Lett. \textbf{71} (1993), 1661--1664.

\bibitem {CLZ}M. Chen, S.-Q. Liu and Y. Zhang, \textit{A 2-component
Generalization of the Camassa--Holm Equation and Its Solutions}, Lett. Math.
Phys. \textbf{75} (2006), 1--15.

\bibitem {C1}A. Constantin, \textit{On the Blow-up Solutions of a Periodic
Shallow Water Equation}, J. Nonlinear Sci. \textbf{10} (2000), 391--399.

\bibitem {CI}A. Constantin and R. Ivanov,\textit{ On an Integrable
Two-component Camassa--Holm Shallow Water System}, Phys. Lett. A \textbf{372}
(2008), 7129--7132.

\bibitem {DXY1}Y.B. Deng, J.L. Xiang and T. Yang, \textit{Blowup Phenomena of
Solutions to Euler-Poisson Equations}, J. Math. Anal. Appl. \textbf{286}
(2003), 295--306.

\bibitem {ELY2}J. Escher, O. Lechtenfeld and Z. Yin, \textit{Well-posedness
and Blow-up Phenomena for the 2-component Camassa--Holm equation}, Discrete
Contin. Dyn. Syst. Ser. A \textbf{19} (2007), 493--513.

\bibitem {GW}P. Goldreich and S. Weber, \textit{Homologously Collapsing
Stellar Cores}, Astrophys. J. \textbf{238 }(1980), 991--997.

\bibitem {GY}C.X. Guan and Z.Y. Yin, \textit{Global Existence and Blow-up
Phenomena for an Integrable Two-component Camassa--Holm Shallow Water System},
J. Differential Equations \textbf{248} (2010), 2003--2014.

\bibitem {G}Z.G. Guo, Blow-up and Global Solutions to a New Integrable Model
with Two Components, J. Math. Anal. Appl. \textbf{372} (2010), 316--327.

\bibitem {GZ}Z.G. Guo and Y. Zhou, On Solutions to a Two-component Generalized
Camassa--Holm System, Stud. Appl. Math. \textbf{124} (2010), 307--322.

\bibitem {H}J. Hale, Theory of Functional Differential Equations, 2nd Edition,
Applied Mathematical Sciences \textbf{3}, Springer-Verlag, New
York-Heidelberg, 1977. x+365 pp.

\bibitem {Li}T. H. Li, Some Special Solutions of the Multidimensional Euler
Equations in $R^{N}$, Comm. Pure Appl. Anal. \textbf{4 }(2005), 757--762.

\bibitem {LW}T. H. Li and D. H. Wang, Blowup Phenomena of Solutions to the
Euler Equations for Compressible Fluid Flow, J. Differential Equations
\textbf{221 }(2006), 91--101.

\bibitem {Liang}Z.L. Liang, \textit{Blowup Phenomena of the compressible Euler
equations}, J. Math. Anal. Appl. \textbf{379 }(2010), 506--510.

\bibitem {M1}T. Makino, \textit{Blowing up Solutions of the Euler-Poission
Equation for the Evolution of the Gaseous Stars}, Transport Theory and
Statistical Physics \textbf{21 }(1992), 615--624.

\bibitem {Wa}W. Walter, Ordinary Differential Equations, Translated from the
6th German (1996) edition by Russell Thompson. Graduate Texts in Mathematics
\textbf{182} Readings in Mathematics, Springer-Verlag, New York, 1998. xii+380 pp.

\bibitem {W}G. B. Whitham, Linear and Nonlinear Waves, Wiley-Interscience, New
York-London-Sydney, 1974.

\bibitem {Y1}M. W. Yuen, \textit{Blowup Solutions for a Class of Fluid
Dynamical Equations in }$R^{N}$ , J. Math. Anal. Appl. \textbf{329} (2007), 1064--1079.

\bibitem {Y44}M. W. Yuen, \textit{Analytical Blowup Solutions to the
2-dimensional Isothermal Euler-Poisson Equations of Gaseous Stars}, J. Math.
Anal. Appl. \textbf{341} (2008)\textbf{, }445--456.

\bibitem {Y2}M. W. Yuen, \textit{Analyitcal Solutions to the Navier-Stokes
Equations}, J. Math. Phys., \textbf{49} (2008), 113102, 10pp.

\bibitem {Y3}M. W. Yuen, \textit{Analytical Blowup Solutions to the
Pressureless Navier-Stokes-Poisson Equations with Density-dependent Viscosity
in }$R^{N}$, Nonlinearity \textbf{22 }(2009), 2261--2268.

\bibitem {Yuen 3}M.W. Yuen, \textit{Self-Similar Blowup Solutions to the
2-Component Camassa-Holm Equations, }J. Math. Phys., \textbf{51} (2010)
093524, 14pp.

\bibitem {Yuen 4}M.W. Yuen, \textit{Self-Similar Blowup Solutions to the
2-Component Degasperis-Procesi Shallow Water System}, Pre-print, arXiv:1008.2282
\end{thebibliography}
\end{document}